
\def\bbbq{{\mathchoice
{\setbox0=\hbox {$\displaystyle\rm Q$}\hbox
{\raise0.15\ht0\hbox to0pt{\kern0.4\wd0\vrule height0.8\ht0\hss}\box0}}
{\setbox0=\hbox {$\textstyle\rm Q$}\hbox
{\raise0.15\ht0\hbox to0pt{\kern0.4\wd0\vrule height0.8\ht0\hss}\box0}}
{\setbox0=\hbox {$\scriptstyle\rm Q$}\hbox
{\raise0.15\ht0\hbox to0pt{\kern0.4\wd0\vrule height0.7\ht0\hss}\box0}}
{\setbox0=\hbox {$\scriptscriptstyle\rm Q$}\hbox
{\raise0.15\ht0\hbox to0pt{\kern0.4\wd0\vrule height0.7\ht0\hss}\box0}}
}}


\def\bbbc{{\mathchoice
{\setbox0=\hbox {$\displaystyle\rm C$}\hbox
{\hbox to0pt{\kern0.4\wd0\vrule height0.9\ht0\hss}\box0}}
{\setbox0=\hbox {$\textstyle\rm C$}\hbox
{\hbox to0pt{\kern0.4\wd0\vrule height0.9\ht0\hss}\box0}}
{\setbox0=\hbox {$\scriptstyle\rm C$}\hbox
{\hbox to0pt{\kern0.4\wd0\vrule height0.9\ht0\hss}\box0}}
{\setbox0=\hbox {$\scriptscriptstyle\rm C$}\hbox
{\hbox to0pt{\kern0.4\wd0\vrule height0.9\ht0\hss}\box0}}
}}

\font\fivesans=cmss10 at 5pt
\font\sevensans=cmss10 at 7pt
\font\tensans=cmss10
\newfam\sansfam
\textfont\sansfam=\tensans\scriptfont\sansfam=\sevensans
\scriptscriptfont\sansfam=\fivesans
\def\sans{\fam\sansfam\tensans}
\def\bbbz{{\mathchoice {\hbox{$\sans\textstyle Z\kern-0.4em Z$}}
{\hbox{$\sans\textstyle Z\kern-0.4em Z$}}
{\hbox{$\sans\scriptstyle Z\kern-0.3em Z$}}
{\hbox{$\sans\scriptscriptstyle Z\kern-0.2em Z$}}}}


\font \bigbf=cmbx10 scaled \magstep2

\def\slash#1{#1\kern-0.65em /}
\def\dirac{{\raise0.09em\hbox{/}}\kern-0.58em\partial}
\def\Dirac{{\raise0.09em\hbox{/}}\kern-0.69em D}




\def\tr{{\rm Tr}}
\vglue 1.5cm

\centerline {\bigbf Linear Connections on Matrix Geometries}
\vskip 1.5cm

\centerline {\bf J. Madore, \ T. Masson}
\medskip
\centerline {\it Laboratoire de Physique Th\'eorique et Hautes
Energies\footnote{*}{\it Laboratoire associ\'e au CNRS.}}
\centerline {\it Universit\'e de Paris-Sud, B\^at. 211,  \ F-91405 ORSAY}
\vskip 1cm

\centerline {\bf J. Mourad}
\medskip
\centerline {\it  Laboratoire de Mod\`eles de Physique Math\'ematique}
\centerline {\it Parc de Grandmont, Universit\'e de Tours, \ F-37200 TOURS}

\vskip 2cm
\noindent
{\bf Abstract:} \ A general definition of a linear connection in
noncommutative geometry has been recently proposed. Two examples are
given of linear connections in noncommutative geometries which are based
on matrix algebras. They both possess a unique metric connection.

\vfill
\noindent
LPTHE Orsay 94/96
\medskip
\noindent
November, 1994
\bigskip
\eject

\beginsection 1 Introduction and Motivation

The extension to noncommutative algebras of the notion of a differential
calculus has been given both without (Connes 1986) and with
(Dubois-Violette 1988) use of the derivations of the algebra. A
definition has been given (Chamseddine {\it et al.} 1993) of a possible
noncommutative generalization of a linear connection which uses the
left-module structure of the differential forms. Recently a different
definition has been given (Mourad 1994, Dubois-Violette {\it et al.}
1994) which makes essential use of the full bimodule structure of the
differential forms. We shall use this definition here to consider linear
connections on two examples of noncommutative geometries based on matrix
algebras. Both have a unique linear connection, which is metric and
torsion free. In this respect they are similar to the quantum plane,
which is not based on a finite-dimensional algebra.

The general definition of a linear connection is given in this section
and in Section~2 some basic formulae from matrix geometry are recalled.
In Section~3 we consider an algebra of forms based on derivations and we
show that there is a unique metric linear connections without torsion.
This case is very similar to ordinary differential geometry and the
calculations follow closely those of this section. In Section~4 we
consider a more abstract differential geometry whose differential
calculus is not based on derivations. Here we find that there is a
unique 1-parameter family of connections, which is without torsion. The
condition that the connection be metric fixes the value of the
parameter.

We first recall the definition of a linear connection in commutative
geometry, in a form (Koszul 1960) which allows for a noncommutative
generalization.  Let $V$ be a differential manifold and let
$(\Omega^*(V), d)$ be the ordinary differential calculus on $V$. Let $H$
be a vector bundle over $V$ associated to some principle bundle $P$. Let
${\cal C}(V)$ be the algebra of smooth functions on $V$ and ${\cal H}$
the left ${\cal C}(V)$-module of smooth sections of $H$.

A connection on $P$ is equivalent to a covariant derivative on $H$,
which in turn can be characterized as a linear map
$$
{\cal H} \buildrel D \over \rightarrow
\Omega^1(V) \otimes_{{\cal C}(V)} {\cal H}                       \eqno(1.1)
$$
which satisfies the condition
$$
D (f \psi) =  df \otimes \psi + f D\psi                          \eqno(1.2)
$$
for arbitrary $f \in {\cal C}(V)$ and $\psi \in {\cal H}$.

The definition of a connection as a covariant derivative  has an
immediate extension to noncommutative geometry. Let ${\cal A}$ be an
arbitrary algebra and $(\Omega^*({\cal A}),d)$ a differential calculus
over ${\cal A}$. We shall define in the next section a differential
calculus $(\Omega^*(M_n),d)$ over the matrix algebras $M_n$. One defines
a covariant derivative on a left ${\cal A}$-module ${\cal H}$ as a map
$$
{\cal H} \buildrel D \over \rightarrow
\Omega^1({\cal A}) \otimes {\cal H}                               \eqno(1.3)
$$
which satisfies the condition (1.2) but with $f \in {\cal A}$.

A linear connection on $V$ can be defined as a connection on the
cotangent bundle to $V$. It can be characterized as a linear map
$$
\Omega^1(V) \buildrel D \over \rightarrow
\Omega^1(V) \otimes_{{\cal C}(V)} \Omega^1(V)                     \eqno(1.4)
$$
which satisfies the condition
$$
D (f \xi) =  df \otimes \xi + f D\xi                              \eqno(1.5)
$$
for arbitrary $f \in {\cal C}(V)$ and $\xi \in \Omega^1(V)$.

Suppose, for simplicity that $V$ is parallelizable and choose
$\theta^\alpha$ to be a globally defined moving frame on $V$. The
connection form $\omega^\alpha{}_\beta$ is defined in terms of the
covariant derivative of the moving frame:
$$
D\theta^\alpha = -\omega^\alpha{}_\beta \otimes \theta^\beta.     \eqno(1.6)
$$
Because of (1.5) the covariant derivative $D\xi$ of an arbitrary element
$\xi = \xi_\alpha \theta^\alpha \in \Omega^1(V)$ can be written as
$D\xi = (D\xi_\alpha) \otimes \theta^\alpha$ where
$$
D\xi_\alpha = d\xi_\alpha - \omega^\beta{}_\alpha \xi_\beta.      \eqno(1.7)
$$
Let $\pi$ be the projection of
$\Omega^1(V) \otimes_{{\cal C}(V)} \Omega^1(V)$ onto $\Omega^2(V)$.
The torsion form $\Theta^\alpha$ can be defined as
$$
\Theta^\alpha = (d - \pi  D)\theta^\alpha.                        \eqno(1.8)
$$

The derivative $D_X\xi$ along the vector field $X$,
$$
D_X \xi = i_X D \xi,                                              \eqno(1.9)
$$
is a linear map of $\Omega^1(V)$ into itself. In particular
$D_X \theta^\alpha =  -\omega^\alpha{}_\beta (X)\theta^\beta$. Using
$D_X$ an extension of $D$ can be constructed  to the tensor product
$\Omega^1(V) \otimes_{{\cal C}(V)} \Omega^1(V)$. We define
$$
D_X(\theta^\alpha \otimes \theta^\beta) =
D_X\theta^\alpha\otimes \theta^\beta +
\theta^\alpha \otimes D_X\theta^\beta                             \eqno(1.10)
$$
Now let $\sigma$ be the action on
$\Omega^1(V) \otimes_{{\cal C}(V)} \Omega^1(V)$ defined by the
permutation of two derivations:
$$
\sigma (\xi \otimes \eta) (X,Y) = \xi \otimes \eta (Y,X)          \eqno(1.11)
$$
and define $\sigma_{12} = \sigma \otimes 1$.  Then (1.10) can be rewritten
without explicitly using the vector field as
$$
D(\theta^\alpha \otimes \theta^\beta) =
D\theta^\alpha\otimes \theta^\beta +
\sigma_{12} (\theta^\alpha \otimes D\theta^\beta).                \eqno(1.12)
$$
Define $\pi_{12} = \pi \otimes 1$. If the torsion vanishes one finds that
$$
\pi_{12} D^2 \theta^\alpha = - \Omega^\alpha{}_\beta \otimes \theta^\beta
                                                                  \eqno(1.13)
$$
where $\Omega^\alpha{}_\beta$ is the curvature 2-form. Notice that
the equality
$$
\pi_{12} D^2 (f \theta^\alpha) = f \pi_{12} D^2 \theta^\alpha     \eqno(1.14)
$$
is a consequence of the identity
$$
\pi (\sigma + 1) = 0.                                              \eqno(1.15)
$$

The module $\Omega^1(V)$ has a natural structure as a right
${\cal C}(V)$-module and the corresponding condition equivalent to (1.5)
is determined using the fact that ${\cal C}(V)$ is a commutative algebra:
$$
D (\xi f) =  D (f \xi).                                            \eqno(1.16)
$$
Using $\sigma$ this can also be written in the form
$$
D(\xi f) = \sigma (\xi \otimes df) + (D\xi) f.                     \eqno(1.17)
$$

By extension, a linear connection over a general noncommutative algebra
${\cal A}$ with a  differential calculus $(\Omega^*({\cal A}),d)$ can be
defined as a linear map
$$
\Omega^1({\cal A}) \buildrel D \over \rightarrow
\Omega^1({\cal A}) \otimes_{{\cal A}} \Omega^1({\cal A})          \eqno(1.18)
$$
which satisfies the condition (1.5) for arbitrary
$f \in {\cal A}$ and $\xi \in \Omega^1({\cal A})$.

The module $\Omega^1({\cal A})$ has again a natural structure as a right
${\cal A}$-module but in the noncommutative case it is impossible in
general to consistently impose the condition (1.16) and a substitute
must be found. We consider first the case where the differential
calculus $(\Omega^*({\cal A}),d)$ is defined using the derivations of
${\cal A}$ (Dubois-Violette 1988).  Let $X$ and $Y$ be arbitrary
derivations of ${\cal A}$ and suppose that the transposition $\sigma$ in
(1.11) maps $\Omega^1({\cal A})\otimes_{\cal A} \Omega^1({\cal A})$ into
itself.  Then we propose to define $D(\xi f)$ by the equation (1.17)
(Dubois-Violette \& Michor 1994a,b).  A covariant derivative is a map of
the form (1.18) which satisfies the Leibniz rules (1.5) and (1.17).  The
right Leibniz rule (1.18) can be made more transparent using the
covariant derivative $D_X$ with respect to the derivation $X$.  The two
Leibniz rules can be written as
$$
\matrix{
D_X(f \xi) = (Xf)\xi + f D_X \xi\hfill,                         \cr
D_X(\xi f) = \xi Xf + (D_X \xi) f.\hfill}                       \eqno(1.19)
$$

A metric $g$ on $V$ can be defined as a ${\cal C}(V)$-bilinear,
symmetric map of $\Omega^1(V) \otimes_{{\cal C}(V)} \Omega^1(V)$ into
${\cal C}(V)$.  This definition makes sense if one replaces
${\cal C}(V)$ by an algebra ${\cal A}$ and $\Omega^1(V)$ by a
differential calculus $\Omega^1({\cal A})$ over ${\cal A}$. By analogy
with the commutative case we shall say that the covariant derivative
(1.17) is metric if the following diagram is commutative:
$$
\matrix{
\Omega^1 \otimes_{\cal A} \Omega^1
&\buildrel D \over \longrightarrow
&\Omega^1 \otimes_{\cal A} \Omega^1 \otimes_{\cal A} \Omega^1         \cr
g \downarrow \phantom{g}&&\phantom{1 \otimes g}\downarrow 1 \otimes g  \cr
{\cal A} &\buildrel d \over \longrightarrow &\Omega^1
}                                                              \eqno(1.20)
$$
We have here set $\Omega^1({\cal A}) = \Omega^1$.  In general symmetry
must be defined with respect to the map $\sigma$. By a symmetric metric
then we mean one which satisfies the condition
$$
g  \sigma = g.                                                 \eqno(1.21)
$$

\beginsection 2 Matrix geometries

Noncommutative geometry is based on the fact that one can formulate
(Koszul 1960) much of the ordinary differential geometry of a manifold
in terms of the algebra of smooth functions defined on it. It is
possible to define a finite noncommutative geometry based on derivations
by replacing this algebra by the algebra $M_n$ of $n\times n$ complex
matrices (Dubois-Violette {\it et al.} 1989, 1990). Since $M_n$ is of
finite dimension as a vector space, all calculations reduce to pure
algebra.  Matrix geometry is interesting in being similar is certain
aspects to the ordinary geometry of compact Lie groups; it constitutes a
transition to the more abstract formalism of general noncommutative
geometry (Connes 1986, 1990). Our notation is that of Dubois-Violette
{\it et al.} (1989). See also Madore (1994). In this section we recall
some important formulae.

Let $\lambda_r$, for $1 \leq r\leq n^2-1$, be an anti-hermitian basis of
the Lie algebra of the special unitary group $SU_n$ in $n$ dimensions.
The $\lambda_r$ generate $M_n$ and the derivations
$$
e_r = {\rm ad}\,\lambda_r                                      \eqno(2.1)
$$
form a basis for the Lie algebra of derivations ${\rm Der}(M_n)$ of
$M_n$.

We define $df$ for $f \in M_n$ by
$$
df(e_r) = e_r(f).                                              \eqno(2.2)
$$
In particular
$$
d\lambda^r(e_s) = - C^r{}_{st} \lambda^t.
$$
We raise and lower indices with the Killing metric $g_{rs}$ of $SU_n$.

We define the set of 1-forms $\Omega^1(M_n)$ to be the set of all
elements of the form $fdg$ with $f$ and $g$ in $M_n$.
The set of all differential forms is a differential algebra
$\Omega^*(M_n)$. The couple $(\Omega^*(M_n), d)$ is a differential
calculus over $M_n$.

There is a convenient system of generators of $\Omega^1(M_n)$ as a
left- or right-module completely characterized by the equations
$$
\theta^r(e_s) = \delta^r_s.                                     \eqno(2.3)
$$
The $\theta^r$ are related to the $d\lambda^r$ by the equations
$$
d\lambda^r = C^r{}_{st}\, \lambda^s \theta^t, \qquad
\theta^r = \lambda_s \lambda^r d\lambda^s.                      \eqno(2.4)
$$
The $\theta^r$ satisfy the same structure equations as the components of
the Maurer-Cartan form on the special unitary group $SU_n$:
$$
d\theta^r =  -{1 \over 2} C^r{}_{st} \, \theta^s \theta^t.      \eqno(2.5)
$$
The product on the right-hand side of this formula is the product in
$\Omega^*(M_n)$. We shall refer to the $\theta^r$ as a frame or
Stehbein. If we define $\theta = - \lambda_r \theta^r$ we can write the
differential $df$ of an element $f \in \Omega^0(M_n)$ as a commutator:
$$
df = - [\theta, f].                                             \eqno(2.6)
$$

\beginsection 3 A differential calculus with derivations

{}From (2.5) we see that the linear connection defined by
$$
D\theta^r = - \omega^r{}_s \otimes \theta^s, \qquad
\omega^r{}_s = -{1\over 2}  C^r{}_{st} \,\theta^t               \eqno(3.1)
$$
has vanishing torsion. With this connection the geometry of $M_n$ looks
like the
invariant geometry of the group $SU_n$.

It follows from the antisymmetry of $C^r{}_{st}$ that
$$
D\theta^r = d\theta^r.
$$
Since the elements of the algebra commute with the frame $\theta^r$, we
can define $D$ on all of $\Omega^*(M_n)$ using (1.5). The map $\sigma$
is given by
$$
\sigma (\theta^r \otimes \theta^s) = \theta^s \otimes \theta^r.\eqno(3.2)
$$
It follows that $D$ satisfies also (1.17).

Consider a general covariant derivative. We can suppose it to be of the form
$$
D\theta^r = -  \omega^r{}_{st} \,\theta^s \otimes \theta^t
                                                               \eqno(3.3)
$$
with $\omega^r{}_{st}$ an arbitrary element of $M_n$ for each value of
$(r,s,t)$. Then from (1.5) and (1.17) we find that
$$
0 = D ([f,\theta^r]) = [f, D\theta^r]                           \eqno(3.4)
$$
and so the $\omega^r{}_{st}$ must be all in the center of $M_n$. They
are complex numbers. If we define the torsion as in (1.8) and require
that it vanish then we have
$$
\omega^r{}_{[st]} =  C^r{}_{st}.                                \eqno(3.5)
$$

Define a metric on $M_n$ by the equation
$g(\theta^r \otimes \theta^s) = g^{rs}$. It satisfies the symmetry
condition (1.21). The commutativity of the diagram (1.20) is the formal
analogue of the condition that a connection be metric. If we impose it
we see that
$$
\omega^r{}_{(st)} = 0.                                           \eqno(3.6)
$$
The linear connection (3.1) is the unique torsion-free metric connection
on $\Omega^1(M_n)$.
{}From the formula analogous to (1.13) we find that the curvature 2-form
is given by
$$
\Omega^r{}_s = {1\over 8} C^r{}_{st} C^t{}_{uv} \theta^u \theta^v.
$$
The connection (3.1) has been used (Dubois-Violette {\it et al.} 1989,
Madore 1990, Madore \& Mourad 1993, Madore 1994) in the construction of
noncommutative generalizations of Kaluza-Klein theories. In particular
the Dirac operator has a natural coupling to it, determined by a
correspondence principle.

\beginsection 4 A differential calculus without derivations

Equation (1.17) can be extended in principle to the case of a
differential calculus which is not based on derivations if we postulate
(Mourad 1994) the existence of a map
$$
\Omega^1 \otimes_{\cal A} \Omega^1 \buildrel \sigma \over \longrightarrow
\Omega^1 \otimes_{\cal A} \Omega^1                               \eqno(4.1)
$$
to replace the one defined by (1.11). We define then $D(\xi f)$ by
the equation (1.17) but using (4.1) instead of (1.11). From the identity
$$
D\big((\xi f)g\big) = D\big(\xi (fg)\big)
$$
we find that $\sigma$ be right ${\cal A}$-linear. From  the identity
$$
D\Big(d\big((fg)h\big)\Big) = D\Big(d\big(f(gh)\big)\Big)
$$
we find that $\sigma$ be also left ${\cal A}$-linear (Dubois-Violette
{\it et al.} 1994). In general
$$
\sigma^2 \neq 1.                                                 \eqno(4.2)
$$
The extension of $D$ to $\Omega^1 \otimes \Omega^1$ is given by (1.12)
but with $\sigma$ defined by (4.1).

As an example we shall consider a differential calculus over an algebra
of matrices with a differential defined by a graded commutator (Connes
\& Lott 1990).  Consider the matrix algebra $M_n$ with a $\bbbz_2$
grading. One can define on $M_n$ a graded derivation $\hat d$ by the
formula
$$
\hat d f = - [ \theta , f],                                        \eqno(4.3)
$$
where $\theta$ is an arbitrary anti-hermitian odd element
and the commutator is taken as a graded commutator. We find that
$\hat d\theta = -2\theta^2$ and for any $\alpha \in M_n$,
$$
\hat d^2 \alpha = [ \theta^2, \alpha ].                            \eqno(4.4)
$$
The $\bbbz_2$ grading of $M_n$ can be expressed as the direct sum
$M_n = M_n^+ \oplus M_n^-$ where $M_n^+$
($M_n^-$) are the even (odd) elements of $M_n$.  It can be
induced from a decomposition $\bbbc^n = \bbbc^l \oplus \bbbc^{n-l}$ for
some integer $l$.  The elements of $M_n^+$ are diagonal with
respect to the decomposition; the elements of $M_n^-$ are
off-diagonal.

It is possible to construct over $M_n^+$ a differential algebra
$\Omega^* = \Omega^*(M_n^+)$ (Connes \& Lott 1991).
Let $\Omega^0 = M^+_n$ and let
$\Omega^1 \equiv \overline{d\Omega^0} \subset M^-_n$ be the
$M^+_n$-bimodule generated by the image of $\Omega^0$ in $M^-_n$
under $\hat d$.  Define
$$
\Omega^0 \buildrel d \over \longrightarrow \Omega^1              \eqno(4.5)
$$
using directly (4.3): $d = \hat d$. Let
$\overline{d\Omega^1}$ be the $M^+_n$-module generated by the
image of $\Omega^1$ in $M^+_n$ under $\hat d$.  It would be natural
to try to set $\Omega^2 = \overline{d\Omega^1}$ and define
$$
\Omega^1 \buildrel d \over \longrightarrow \Omega^2              \eqno(4.6)
$$
using once again (4.3). Every element of $\Omega^1$ can be
written as a sum of elements of the form $f_0 \hat d f_1$. If we attempt
to define an application (4.6) using again directly (4.3),
$$
d (f_0 \hat d f_1) = \hat df_0 \hat df_1 + f_0 \hat d^2 f_1,      \eqno(4.7)
$$
then we see that in general $d^2$ does not vanish. To remedy this
problem we eliminate simply the unwanted terms. Let  ${\rm Im}\,\hat d^2$
be the submodule of $\overline{d\Omega^1}$ consisting of those elements
which contain a factor which is the image of $\hat d^2$ and define
$\Omega^2$ by
$$
\Omega^2 = \overline{d\Omega^1} / {\rm Im}\,\hat d^2.\eqno(4.8)
$$
Then by construction the second term on the right-hand side of (4.7)
vanishes as an element of $\Omega^2$ and we have a well defined map
(4.6) with $d^2 = 0$.  This procedure can be continued to arbitrary
order by iteration.  For each $p \geq 2$ we let ${\rm Im}\,\hat d^2$ be
the submodule of $\overline{d\Omega^{p-1}}$ defined as above and
we define $\Omega^p$ by
$$
\Omega^p = \overline{d\Omega^{p-1}} / {\rm Im}\,\hat d^2.
                                                                  \eqno(4.9)
$$
Since $\Omega^p \Omega^q \subset \Omega^{p+q}$ the complex
$\Omega^*$ is a differential algebra.  The $\Omega^p$ need not
vanish for large values of $p$. In fact if $\theta^2 \propto 1$ we see
that $\hat d^2 = 0$ and the sequence defined by (4.9) never
stops.  However $\Omega^p \subseteq M^+_n (M^-_n)$ for
$p$ even (odd) and so it stabilizes for large $p$.

We shall consider in some detail the case $n=3$ with the grading
defined by the decomposition $\bbbc^3 = \bbbc^2 \oplus \bbbc$.
The most general possible form for $\theta$ is
$$
\theta = \eta_1 - \eta_1^*                                        \eqno(4.10)
$$
where
$$
\eta_1 = \pmatrix{0  &  0  &     a  \cr
                  0  &  0  &     b  \cr
                  0  &  0  &     0}.                          \eqno(4.11)
$$
Without loss of generality we can choose the euclidean 2-vector
$\eta_{1i}$ of unit length. The general construction yields
$\Omega^0 = M_3^+ = M_2 \times M_1$ and $\Omega^1 = M_3^-$
but after that the quotient by elements of the form ${\rm Im}\,\hat d^2$
reduces the dimension. One finds $\Omega^2 = M_1$ and
$\Omega^p = 0$ for $p\geq 3$. Let $e$ be the unit of $M_1$. It generates
$\Omega^2$ and can also be considered as an element of $\Omega^0$.

To form a basis for $\Omega^1$ we must introduce a second matrix
$\eta_2$.  It is convenient to choose it of the same form as $\eta_1$.
We have then in $\Omega^2$ the identity
$$
\eta_i \eta^*_j = 0.
$$
We shall further impose that
$$
\eta^*_i \eta_j = \delta_{ij} e.                                \eqno(4.12)
$$
It follows that
$$
d\eta_1 = e,   \qquad  d\eta_2 = 0.
$$
We can uniquely fix $\eta_2$ by requiring that there be a unitary
element $u \in M_2 \subset M_3^+$ which exchanges $\eta_1$ and $\eta_2$:
$$
\eta_2 = u \eta_1, \qquad \eta_1 = - u \eta_2.                  \eqno(4.13)
$$
We have also
$$
\eta_2 u = 0, \qquad \eta_1 u = 0.                              \eqno(4.14)
$$

The vector space of 1-forms is of dimension 4 over the complex numbers.
The dimension of $\Omega^1 \otimes_\bbbc\Omega^1$ is equal to 16 but the
dimension of the tensor product $\Omega^1 \otimes_{M_3^+}\Omega^1$ is
equal to 5.  One finds in fact over $M_3^+$ the relations
$$
\eqalign{
&\eta_i \otimes \eta_j = 0,                         \qquad
 \eta^*_i \otimes \eta^*_j = 0,                     \qquad\cr
&\eta^*_2 \otimes \eta_1 = 0,                       \qquad
 \eta^*_1 \otimes \eta_2 = 0,                       \qquad
\eta^*_2 \otimes \eta_2 =  \eta^*_1 \otimes \eta_1.}            \eqno(4.15)
$$
which leave
$$
\eta_{ij} = \eta_i \otimes \eta^*_j, \qquad
\zeta = \eta^*_1 \otimes \eta_1                                 \eqno(4.16)
$$
as independent basis elements. We can make therefore the identification
$$
\Omega^1 \otimes_{M_3^+}\Omega^1 = M_3^+ = \Omega^0.            \eqno(4.17)
$$

To define a covariant derivative we must first introduce the map
$\sigma$ of (4.1). Because of the left and right $M_3^+$-linearity
the map $\sigma$ is entirely determined by its action on $\zeta$ and,
for example, $\eta_{11}$:
$$
\eqalign
{
&\sigma(\eta_{11}) = \sum_{ij} a_{ij} \eta_{ij} + a \zeta, \cr
&\sigma(\zeta) = \sum_{ij} b_{ij} \eta_{ij} + b \zeta.
}
$$
If we multiply both sides of the second equation by $u$ we find that
$b_{ij} = 0$; if we multiply both sides of the first equation by $u^2$
we find that $a = 0$. Let $v$ be a matrix such that $v\eta_1 = \eta_1$
and $v\eta_2 \neq \eta_2$. From the conditions of left and right
linearity we have the equations
$$
\sigma(\eta_{11}) = v \sigma(\eta_{11})
                  = \sum_{ij} a_{ij} v \eta_{ij},                 \qquad
\sigma(\eta_{11}) = \sigma(\eta_{11})v^*
                  = \sum_{ij} a_{ij} \eta_{ij}v^*,
$$
from which we conclude that
$$
a_{11} = \mu, \qquad a_{12} = a_{21} = a_{22} = 0,
$$
where $\mu$ is an arbitrary complex number.  If we impose the condition
(1.15) we find that $1 + b = 0$. So $\sigma$ is given by
$$
\sigma(\eta_{11}) = \mu \eta_{11}, \qquad
\sigma(\zeta) = - \zeta.                                       \eqno(4.18)
$$

The Hecke relation
$$
(\sigma + 1)(\sigma - \mu) = 0
$$
is satisfied. Suppose that $\mu \neq -1$ and define $\Lambda^*$ ($S^*$)
to be the quotient of the tensor algebra by the ideal generated by the
eigenvectors of $\mu$ ($-1$). As a complex vector space $\Lambda^*$ is
of dimension 10.  The map $\zeta \mapsto e$ induces an isomorphism of
$\Lambda^*$ with $\Omega^*$.  As a complex vector space $S^*$ is of
dimension 13. It is an unusual fact that it is of finite dimension.  If
$\mu = -1$ then $\sigma = -1$ also and (1.15) is trivially satisfied.
In this case it is natural to define $\Lambda^*$ to be the entire tensor
algebra.  On the universal differential calculus the projection $\pi$ of
(1.15) is the identity and $\sigma$ must be equal to $-1$.

The covariant derivative of $\eta_i$ must be of the form
$$
D\eta_i = \sum_{jk} c_{ijk} \eta_{jk} + c_i \zeta.
$$
The exterior derivative of $u$ is given by
$$
du = \eta_2 - \eta^*_2.                                        \eqno(4.19)
$$
{}From (1.5) we find then that $D$ must satisfy the constraints
$$
D\eta_1 = \zeta - u D\eta_2, \qquad
D\eta_2 =  u D\eta_1                                           \eqno(4.20)
$$
and therefore that $c_1 = 1$, $c_2 = 0$ and $c_{2ij}$ is determined in
terms of $c_{1ij}$:
$$
c_{211} = - c_{121}, \quad     c_{212} = - c_{122}, \quad
c_{221} =   c_{111}, \quad     c_{222} =   c_{112}. \quad
$$

{}From the condition (1.17) one finds the additional constraints
$$
(D\eta_1) u - \sigma(\eta_{12}) = 0. \qquad
(D\eta_2) u - \sigma(\eta_{22}) = 0,
$$
which both imply that
$$
c_{111} = - \mu,  \qquad c_{112} = 0,  \qquad
c_{121} = 0.      \qquad c_{122} = 0                           \eqno(4.21)
$$
The covariant derivative is uniquely defined then in terms of $\sigma$:
$$
D\eta_1 = -\mu \eta_{11} + \zeta, \qquad
D\eta_2 = -\mu \eta_{21}.                                      \eqno(4.22)
$$
The lack of symmetry is due to the fact that the form $\theta$ which
determines the exterior derivative is defined in terms of $\eta_1$.  The
torsion vanishes. Recently (Dubois-Violette {\it et al.} 1994) the
quantum plane has been shown to possess a 1-parameter family of covariant
derivatives, which also are torsion free. If one takes the covariant
derivative of the identity $\theta e = \eta_1$ and its adjoint one finds
that
$$
D\eta^*_1 = - \eta_{11} - \zeta, \qquad
D\eta^*_2 = - \eta_{12},                                       \eqno(4.23)
$$
and therefore that
$$
D\theta = (\sigma -1 ) \, \theta \otimes \theta.               \eqno(4.24)
$$
One finds also that
$$
D(de) = (1 + \mu) \eta_{11}, \qquad D(du) = \eta_{12} - \mu \eta_{21}.
                                                               \eqno(4.25)
$$
Using the identification (4.17) one sees that
$$
\eqalign{
&\hat d \eta_1 = -\theta \otimes \eta_1 - \eta_1 \otimes \theta
              = \eta_{11} + \zeta,                             \cr
&\hat d \eta^*_1 =  -\theta \otimes \eta^*_1 - \eta^*_1 \otimes \theta
                = - \eta_{11} - \zeta
}                                                               \eqno(4.26)
$$
Therefore when $\mu = - 1$ one can identify $D$ with $\hat d$.

Let $g$ be a metric and set
$$
h_{ij} = g(\eta_{ij}), \qquad h = g(\zeta).
                                                                \eqno(4.27)
$$
If we suppose that the metric is bilinear then $h_{ij}$ is given
in terms of $h_{11}$. For example
$$
h_{21} = u h_{11}.                                              \eqno(4.28)
$$
The condition that the connection be metric compatible is expressed by
the equations
$$
\eqalign{
&dh_{11} =  - \mu \eta_1 h +     \eta^*_1 h_{11},               \cr
&dh      =  -     \eta_1 h + \mu \eta^*_1 h_{11}.
}                                                               \eqno(4.29)
$$
This equation has no solutions unless $\mu^2 = 1$. If $\mu = 1$ then to
within an overall scale the unique bilinear metric is given by
$$
h_{ij} = \eta_i \eta^*_j, \qquad h = - e,                       \eqno(4.30)
$$
where the right-hand sides are considered as elements of $M^+_3$.
Therefore $h_{ij}$ ($h$) takes its values in the $M_2$ ($M_1$) factor of
$M^+_3$.  From (4.18) we see that the metric (4.30) is not symmetric.
With the normalization we have chosen the frames $\eta_i$ have unit norm
with respect to the metric:
$$
\tr (h_{ij}) = \delta_{ij}.                                     \eqno(4.31)
$$

The curvature can be defined by a formula analogous to (1.13).
Using (1.12) we find
$$
\matrix{
D\eta_{11} = \zeta \otimes \eta^*_1 - \mu \eta_{11} \otimes \eta_1,\hfill
&D\eta_{12} = \zeta \otimes \eta^*_2,                           \hfill\cr
D\eta_{21} = - \mu \eta_{21} \otimes \eta_1,                    \hfill
&D\zeta = \mu\zeta \otimes \eta^*_1 - \eta_{11} \otimes \eta_1, \hfill
}                                                               \eqno(4.32)
$$
from which we conclude that
$$
\matrix{
D^2 \eta_1 = (\mu^2 - 1) \eta_{11} \otimes \eta_1,      \hfill
&D^2 \eta_2 = \mu^2 \eta_{21} \otimes \eta_1,           \hfill\cr
D^2 \eta^*_1 = (\mu + 1) (\eta_{11} \otimes \eta_1
             - \zeta \otimes \eta^*_1),                 \hfill
&D^2 \eta^*_2 =  - \zeta \otimes \eta^*_2.              \hfill
}                                                                \eqno(4.33)
$$
The curvature is given by the projection of $D^2 \eta_1$ onto
$\Omega^2 \otimes_{M^+_3} \Omega^1$:
$$
\matrix{
\pi_{12} D^2 \eta_1 = 0,                                \hfill
&\pi_{12} D^2 \eta_2 = 0,                               \hfill\cr
\pi_{12} D^2 \eta^*_1 = -(\mu + 1) e \otimes \eta^*_1,  \hfill
&\pi_{12} D^2 \eta^*_2 = - e \otimes \eta^*_2.          \hfill
}                                                                \eqno(4.34)
$$
Although by construction the operator $\pi_{12} D^2$ is left linear it
is not right linear.

For no value of $\mu$ does the curvature vanish. However the analogue of
the square of the curvature tensor does vanishes. In fact, the tensor
product of the curvature tensor with itself vanishes identically.
There are 4 different frames corresponding to the 4 different ways of
choosing $\eta_i$ and $\eta^*_i$ as generators of $\Omega^1$. The action
of the matrix $u$ which takes one into the other is a change of frame.
Since $h_{ij}$ is not proportional to the identity matrix, the frames
cannot be considered as the analogues of orthonormal frames and since
$h_{22} \neq h_{11}$ the change of frame $u$ cannot be considered as
`orthonormal'. If we define
$$
\pi_{12} D^2 \eta_i = - R_{(i)} e \otimes \eta_i,          \qquad
\pi_{12} D^2 \eta^*_i = - \bar R_{(i)} e \otimes \eta^*_i,       \eqno(4.35)
$$
we see that $R$ vanishes in all frames and that
$$
\bar R_{(1)} = \mu + 1,   \qquad \bar R_{(2)} = 1.               \eqno(4.36)
$$
When we take $\eta^*_1$ into $\eta^*_2$ by the right action of $u$ we
change the value of $\bar R$ from $\mu + 1$ to $1$.

Since $\eta_1 = \theta e$ and $\eta^*_1 = - e \theta$ we could also
choose $\theta$ as frame. If we rewrite Equation~(4.35) is this frame,
$$
\pi_{12} D^2 \theta = - R_{(\theta)} e \otimes \theta,           \eqno(4.37)
$$
we see that the component of the curvature can be defined by one matrix,
proportional to the identity matrix, given by
$$
 R_{(\theta)} = \mu + 1.                                         \eqno(4.38)
$$

The analogue of the Ricci tensor would be obtained by using the metric to
`contract two indices' of the curvature tensor. We can do this here
if we identify $\Omega^2$ with the vector space $\Lambda^2$.
We can define a left-linear map Ric of $\Omega^1$ into itself by
$$
{\rm Ric} (\xi) = - (1 \otimes g) \pi_{12} D^2 \xi.              \eqno(4.39)
$$
{}From, for example, the identity
$$
(1 \otimes g) (\zeta \otimes \eta^*_1)
= \eta^*_1 h_{11} = \eta^*_1                                     \eqno(4.40)
$$
ones sees that Ric is given by the equations
$$
{\rm Ric} (\eta_i) = R_{(i)} \eta_i,                 \quad
{\rm Ric} (\eta^*_i) = \bar R_{(i)} \eta^*_i.                    \eqno(4.41)
$$
The geometry is therefore not `Ricci-flat'. There is no analogue of the
Ricci scalar.

There does not seem to be any way to construct a frame-independent
quantity so the best we can do is declare $\theta$ to be a preferred
frame and consider the component (4.38) of the curvature in this frame
as the curvature of the geometry of $M^+_3$. If we require that it be
metric the connection is unique and any action would yield it as
extremal. We could on the other hand consider $\mu$ as an unknown
parameter and chose as action
$$
\tr (R_{(\theta)}^2) = 3(\mu + 1)^2.                             \eqno(4.42)
$$
The action has then a minimal which corresponds to a connection which is
not metric and whose curvature component vanishes in the frame $\theta$.

Additional structure could be put on the algebra $M^+_3$. For example
one could replace the $M_2$ component with the algebra of quaternions or
require that the matrices be hermitian. In the latter case the two
possible frames are $\eta_1 + \eta^*_1$ and $\eta_2 + \eta^*_2$. They
yield each one curvature component whose values are given by (4.36).

We mentioned that the Dirac operator has a natural coupling to the
geometry of the previous section. There is also a generalized
correspondence principle which can be used as a guide in introducing the
Dirac operator coupled to the geometry of the quantum plane. The
coupling of the Dirac operator to the geometry considered here is
however more problematic. There is no possible correspondence principle
since the geometry is not a deformation of a commutative geometry. It is
natural to require that a spinor be an element of a left $M_3$-module
and that the Dirac operator be an hermitian element of $M^-_3$ but
otherwise there is no restriction. In ordinary geometry the exterior
derivative can be identified with the commutator of the Dirac operator
(Connes 1986) and this has been used as motivation for proposing
$i\theta$ as the Dirac operator in the present case, without any
consideration of curvature (Connes \& Lott 1990).  This or any other
element of $M^-_3$ could be considered as automatically coupled to the
curvature since there is a unique metric connection.

Since we have a differential calculus we have an associated cohomology.
By definition $H^0 = M^+_3$ and $H^p = 0$ for $p \geq 3$. The unique
2-cocyle $e$ is the coboundary of $\eta_1$ and so $H^2 = 0$. The vector
space $Z^1$ of 1-cocycles is of complex dimension 2, generated by
$\eta_2$ and $\eta^*_2$. From (4.19) we see that $\eta_2 - \eta^*_2$ is
a coboundary; it is easy to verify that so also is $\eta_2 + \eta^*_2$.
Therefore $H^1 =0$ and the cohomology is trivial.

\parskip 7pt plus 1pt
\parindent=0cm
{\it Acknowledgment:}\ The authors would like to thank M.
Dubois-Violette for enlightening comments.
\vfill\eject

\beginsection References

Chamseddine A.H., Felder G., Fr\"ohlich J. 1993, {\it Gravity in
Non-Commutative Geometry}, Commun. Math. Phys. {\bf 155} 205.

Connes A. 1986, {\it Non-Commutative Differential Geometry}, Publications
of the Inst. des Hautes Etudes Scientifique. {\bf 62} 257.

--- 1990, {\it G\'eom\'etrie noncommutative}, InterEditions, Paris.

Connes A., Lott J. 1990, {\it Particle Models and Noncommutative Geometry},
in `Recent Advances in Field Theory', Nucl. Phys. Proc. Suppl. {\bf B18} 29.

--- 1991, {\it The metric aspect of non-commutative
geometry}, Proceedings of the Carg\`ese Summer School, (to appear).

Dubois-Violette M. 1988, {\it D\'erivations et calcul diff\'erentiel
non-commutatif}, C. R. Acad. Sci. Paris {\bf 307} S\'erie I 403.

Dubois-Violette M., Kerner R., Madore J. 1989, {\it Gauge bosons in a
noncommutative geometry}, Phys. Lett. {\bf B217} 485; {\it Classical
bosons in a noncommutative geometry}, Class. Quant. Grav. {\bf 6} 1709.

--- 1990, {\it Noncommutative differential geometry of matrix algebras},
J. Math. Phys. {\bf 31} 316.

Dubois-Violette M., Madore J., Masson T., Mourad J. 1994, {\it Linear
Connections on the Quantum Plane}, Preprint LPTHE Orsay 94/94.

Dubois-Violette M., Michor P. 1994a, {\it D\'erivations et calcul
diff\'erentiel non-commuta\-tif II}, C. R. Acad. Sci. Paris {\bf 319}
S\'erie I 927.

Dubois-Violette M., Michor P. 1994b, Private communication.

Koszul J.L. 1960, {\it Lectures on Fibre Bundles and Differential Geometry},
Tata Institute of Fundamental Research, Bombay.

Madore J. 1990, {\it Modification of Kaluza-Klein Theory}, Phys. Rev.
{\bf D41} 3709.

Madore J. 1994, {\it An Introduction to Noncommutative Differential Geometry
and its Physical Applications}, Cambridge University Press (to appear).

Madore J., Mourad J. 1993, {\it Algebraic-Kaluza-Klein Cosmology}, Class.
Quant. Grav. {\bf 10} 2157.

Mourad. J. 1994, {\it Linear Connections in Non-Commutative Geometry},
Univ. of Tour Preprint.

\bye